\documentstyle[preprint,aps]{revtex}
\begin{document}
\baselineskip=24pt 
\title {Crowd effects and volatility in a competitive market}  
\author{Neil F. Johnson$^{*}$, Michael Hart} 
\address {Physics Department, Oxford  University, Parks Road,
Oxford, OX1 3PU, England}
\author{P. M. Hui}
\address {Department of Physics, The Chinese University of Hong
Kong, Shatin, Hong Kong}  
\maketitle
\vskip\baselineskip 
\begin{abstract}  We present analytic and numerical results for
two models, namely the minority model and the bar-attendance
model, which offer simple paradigms for a competitive
marketplace. Both models feature heterogeneous agents with
bounded rationality who act using inductive reasoning.  We find
that the effects of crowding are crucial to the understanding
of the macroscopic fluctuations, or `volatility', in the
resulting dynamics of  these systems.

\vskip\baselineskip

\vskip0.5cm
\noindent {\bf $^*$}  n.johnson@physics.oxford.ac.uk  

\end{abstract}

%\narrowtext

\newpage

\noindent {\bf 1. Introduction}

There is much attention now being paid in various disciplines to
the topic of complex adaptive systems
\cite{Casti,bak,physics,Palermo}. In an economics context, such
systems may provide invaluable insight since they can avoid the
standard economic assumptions of equilibrium based on rational
behaviour by agents. It is now recognised among
economists that realistic models of economic markets should
include the effects of heterogeneous agents with bounded
rationality acting via inductive reasoning.  In finance theory,
a central problem is to understand the microscopic factors which
give rise to macroscopic fluctuations, or `volatility', in
financial markets. It would be fascinating if complex adaptive
systems could be used to explain such global market properties.

Here we present analytic and numerical results for two models
which offer simple paradigms for a competitive marketplace
containing competing agents.  We show that the effects of
crowding are crucial for understanding the macroscopic
fluctuations in the resulting dynamics of these systems.  For
the minority model, introduced by Zhang and co-workers
\cite{Challet,Savit}, we find that an analytic model
incorporating `crowd' and `anticrowd' effects can explain the
numerically-obtained volatility over a wide range of parameter
space.  For the bar-attendance model, introduced by Arthur
\cite{Arthur}, we also find that crowd effects are essential for
understanding the volatility. 
\vskip\baselineskip

\noindent {\bf 2. Minority Model}

The minority model was introduced by Challet and Zhang
\cite{Challet} and was also discussed by Savit {\em et al}.
\cite{Savit} and de Cara {\em et al} \cite{Cara}. The basic
model takes the form of a repeated game as follows.  Consider
an odd number of agents
$N$ who must choose whether to be in room `0' or room `1'. 
After every agent has independently chosen a room, the winners
are those in the minority room, i.e.,  the room with fewer
agents. The `output' is a single binary digit for each
time-step: 0 represents room 0 winning, 1 represents room 1
winning.  This output is made available to all agents, and is
the {\em only} information they can use to decide which room to
choose in subsequent time-steps.  Given that the agents are of
limited yet similar capabilities, we assign to each agent a
`brain-size' $m$; this is the length of the past history
bit-string that an agent can use when making its next decision.
Consider $m=2$; the possible
$m=2$ history bit-strings are 00, 01, 10, 11. When faced with
any one of these $2^m=4$ histories, the agent can make one of
two decisions, 0 or 1. Hence, there are
$2^{2^m}=16$ possible strategies which define the decisions in
response to all possible $m=2$ history bit-strings. Each
strategy can thus be represented by a string of 4 bits
[$ijk\ell$] with
$i,j,k,\ell = 0$ or $1$ corresponding to the decisions based on
the histories 00, 01, 10, 11, respectively.   For example,
strategy [0000] corresponds to deciding to pick room 0
irrespective of the $m=2$ history bit-string. [1111] corresponds
to  deciding to pick room 1 irrespective of the $m=2$ history
bit-string. [1010] corresponds to  deciding to pick room 1 given
the histories 00 or 10, and pick room 0 given the histories 01
or 11. The agents randomly pick $s$ strategies at the beginning
of the game. After each turn of the game, the agent assigns one
(virtual) point to each of his strategies which would have
predicted the correct outcome. In addition the agent gets
awarded one (real) point if he is successful. At each turn of
the game, the agent uses whichever is the most successful
strategy among the
$s$ strategies in his possession, i.e., he chooses the one that
has gained most (virtual) points. 

The important feature of both the minority model, and the
bar-attendance model described later, is that the success of any
particular strategy is generally short-lived. If all the agents
begin to use similar strategies, and hence choose the same room,
such a strategy ceases to be profitable and is hence dropped.
Hence there is no best strategy for predicting the market for
all times. References
\cite{Challet} and
\cite{Savit} provide various numerical results for the minority
model.  The main result which emerges from the numerical
simulations
\cite{Challet,Savit} of the minority model concerns the
volatility (i.e., standard deviation) of the time-series $x(t)$
corresponding to the number of agents attending a given room,
say room 0, at each time step, with each time step taken to be
a turn in the game.  When the number of strategies per agent
$s$ is small, the volatility $\sigma$ exhibits a pronounced
minimum as a function of the brain-size $m$. Around this
minimum, the volatility
$\sigma$ is substantially {\em smaller} than the value 
obtained for the case where each agent makes his decision
by tossing a coin
\cite{Challet,Savit}.  A complete analytic theory describing
$\sigma$ as a function of $m$ for arbitrary $N$ and $s$ has,
however, been lacking.
\vskip\baselineskip

\noindent {\bf 3. Effects of Crowds and Anticrowds}

Here we propose an analytic theory which describes the
volatility for arbitrary $m$, $N$ and
$s$. We first present the basic idea, before proceeding to
explicit expressions. Consider the oversimplified case of $N$
independent agents each deciding whether or not to attend room
0 by tossing a coin. Using standard random-walk results, the
total variance
$\sigma^2$ of attendance in room 0 is given by the sum of the
variances produced by the $N$ independent agents: 
\begin{equation}
\sigma^2 = {\sum_{i=1}^N} \sigma_i^2
\end{equation} where $\sigma_i^2 = \frac{1}{2}  . 
(1-\frac{1}{2})=\frac{1}{4}$. Hence $\sigma^2=\frac{N}{4}$.  
However in reality on any given turn of the minority game, there
are a number of agents using the same, or similar, strategies.
Consider the subset of agents
$n_i$ using a particular strategy
$i$. Although there is no information available to a given
agent about other individual agents nor is any direct
communication allowed between agents, this subset
$n_i$ will all act in the same way, i.e., they all go to either
room 0 or 1 and hence constitute a {\em crowd}.  Since the
corresponding random-walk `step-size' has become $n_i$,  one
might think that
$\sigma_i^2$ should be given by
$ \frac{1}{4}  n_i^2 $. Given that there is no a priori best
strategy, however, it is important to realize that there may
also be a subset of agents $n_{\bar i}$ who are using the
opposite, or at least very dissimilar, strategies to the first
subset $n_i$. We call this second subgroup the {\em anticrowd},
and the strategy $\bar i$ that they use  is anti-correlated to
strategy $i$ (e.g. if $i$ is [0000] then $\bar i$ is [1111])
\cite{Challet}. The {\em anticrowd} chooses the opposite room
to the {\em crowd} and hence behaves as a crowd itself.  Over
the timescale during which the two opposing strategies are
being played, the fluctuations of attendance in room 0 are 
determined only by the net crowd-size 
$N_i=n_i-n_{\bar i}$. Hence $\sigma_i^2$ should instead be
given by
$\frac{1}{4}
 N_i^2 $.  

Suppose strategy $i^*$ is the highest scoring at a particular
moment in the game: the anti-correlated strategy $\bar{i^*}$ is
therefore the lowest scoring at that same moment.   In the
limit of small brain-size $m$, the size of the strategy space
is small.  For most values of $s$, it follows that even if an
agent picks $\bar{i^*}$ among his $s$ strategies, he will not
have to use it since he will most likely have a high scoring
strategy in his toolbag.  In practice if $m$ is small, the
required
$s$  can be relatively small since a modest value of $s$ is
sufficient for  each agent to carry a considerable fraction of
all possible  strategies.   Therefore, many agents will choose
to use either
$i^*$ itself (if they hold it among their $s$ strategies) or a
similar one.  Very few agents will have such a poor set of $s$
strategies that they are forced to use a strategy similar to
$\bar {i^*}$.   In this regime there are practically no
anticrowds, hence the crowds dominate. Therefore
$N_i \sim N \delta_{ii^*}$ yielding $\sigma_i^2 \sim
\frac{N^2}{4}
\delta_{ii^*}$; the resulting volatility $\sigma^2 \sim
\frac{N^2}{4}$ is larger than the independent agent limit of
$\frac{N}{4}$ and is therefore consistent with the numerical
simulations \cite{Challet,Savit}. In the opposite limit of large
brain-size $m$, the strategy space is very large, hence agents
will have a low chance of holding, and hence playing, the same
strategy.  In addition, even if an agent has $s$ low-scoring
strategies, the probability of his best strategy being strictly
anticorrelated to another agent's best strategy (hence forming a
crowd-anticrowd pair) is small.  All the crowds and
anticrowds are of size $0$ or $1$ implying that  the crowds and
anticrowds have effectively disappeared and the agents act
independently.  We thus have 
$N_i=0$ or $1$ with $\sum_i N_i\sim N$.   The volatility is
then given by
$\sigma^2\sim\frac{N}{4}$ which is again  consistent with the
numerical simulations
\cite{Challet,Savit}. In the intermediate $m$ region where the
numerical minimum exists for small $s$, the size of the
strategy space is relatively large so that some agents may get
stuck with $s$ strategies which are all low scoring.  They
can hence form anticrowds.   The presence of finite-size
anticrowds implies that
$\sum_i N_i<N$. Considering the extreme case where the crowd and
anticrowd are of similar size, we have
$N_i
\sim 0$ and hence $\sigma_i^2 \sim 0$. The  volatility is
therefore small  ($\sigma \sim 0$) which is again consistent
with the numerical results.  For a fixed value of $m$, this
regime of small volatility will arise for small
$s$ since in this case the number of strategies available to
each agent is small, hence some of the agents may indeed  be
forced to use  a strategy which is little better than the
poorly-performing $\bar {i^*}$.  In other words, the
cancellation effect of the crowd and anticrowd becomes most
effective in this intermediate region.  It is this interplay
between the size of the strategy space and the probabilities 
of agents forming crowds and anticrowds which gives rise to the
rich and non-trivial  behaviour of the volatility.  

Analytic expressions for the crowd and anticrowd sizes can be
obtained in various ways, with varying degrees of accuracy.
Here we present one such approach, and show that it yields good
quantitative agreement with the numerical results. We need to
calculate the number of agents who are using a given strategy
$i$ or one which is similar, such that they are acting
identically in the majority of turns. We can immediately
exploit the results of Ref. \cite{Challet} concerning the
reduced strategy space. In particular, given a strategy $i$,
the only strategies that are significantly different from $i$
are the anti-correlated strategy
$\bar {i^*}$ plus the uncorrelated strategies \cite{Challet}.
For example for $m=2$, given a strategy $i\equiv [0000]$ the 
strategies which will yield significantly different actions
from $i$ are $\bar{i}
\equiv [1111]$, plus the uncorrelated strategies [1100],
[1010], [1001], [0110], [0101], [0011]. Although  the full
strategy space contains
$2^{2^m}=16$ strategies,  the reduced strategy space only
contains $2 \cdot 2^m=8$. Henceforth we only need to consider
the reduced strategy space -- in agreement with Ref. \cite
{Challet}, we have found that the numerical results for the
volatility are quantitatively very similar for both the full
strategy space and the reduced strategy space
\cite{Challet}. 

The number of strategies in the reduced strategy space $V_m$ is
given by
$a=2\cdot 2^m$. For $s=1$ strategy per agent, the probability
that a given strategy in
$V_m$ will be picked by a given agent at the start of the game
is
$\frac{1}{a}$. Consider a given moment in the game. We can rank
all the strategies as best, 2nd best, 3rd best, etc at that
moment. Since agents will always play their best strategy, the
probability that the current best  strategy is being used by a
given agent is given by  the probability that the agent actually
has this strategy. Similarly, the probability that the 2nd
best  strategy is being used by a given agent is equal to the
probability that the agent has this strategy but does {\em not}
possess the best strategy. Since repetition of picked
strategies is allowed during the initial picking process, it is
straightforward to show that for general
$s$ the mean number of agents $y_{r}$ using the $r$-th best
strategy is given by
\begin{equation} y_r=N\bigg(\bigg[1 - \frac{(r-1)}{a}\bigg]^s -
\bigg[1-\frac{r}{a}\bigg]^s\bigg)
\ .
\end{equation} Notice that $\sum_{r=1}^{r=a} y_r =N$ as
required. 

Next we consider the probability $P$ that given a
particular strategy is being used, then its anticorrelated
strategy is also
being used.   Consider the list of all
$2 \cdot 2^m$ strategies in order of the number of virtual
points they have collected. The $i$-th strategy and the $[2
\cdot 2^m +1 -i]$-th strategy will be anticorrelated.   For
large values of $m$  and hence large reduced strategy space,
$P$ will be negligible since  it is highly unlikely 
that if a strategy is used then its anticorrelated strategy is
also used.  It is only  when the total number of strategies in
play, which is of order
$N$,  greatly exceeds the number of strategies in the reduced
strategy space that $P$ will be of order unity.  We can thus
approximate
$P$  by 
$P(p)=p$ for $p<1$ and $P(p)=1$ for $p>1$, where $p=N /(2 \cdot
2^m)$.  Although the form for $P(p)$ can be made more accurate, 
the present expression is reasonable since there are only of
order
$N$ strategies out of a possible maximum of $2 \cdot 2^m$ which
can actually be in play at any one time. Hence, as expected,
$P(p)$ is zero when
$N<<2\cdot 2^m$ and unity when $N>>2\cdot 2^m$. We have checked
that our analytic results for the volatility are fairly
insensitive to the precise form of $P(p)$ as long as $P(p)$
satisifes the constraints
$P(0)=0$ and $P(p>>1)=1$.

Equation (2) gives $y_{r}$ in terms of $m$, $s$, and $N$.   In
satisfying $\sum_{r=1}^{r=a} y_r =N$, we should in practice 
take into account the fact that agents exist only as
integer values. Hence we should only include $R$ terms
in this sum, subject to the condition that the
partial sum equals
$N$ {\em after} the quantities $y_r$  have been rounded to the
nearest integer.  In addition we choose to round any $y_r$'s
which are less than one, up to one if $r\leq R$ such that the
first
$R$ terms are all non-zero. There are hence only $R$ different
strategies in play; note that $R\leq 2 \cdot 2^m$ and
$R\leq N$. The probability
$P(p)$ is, to a reasonable approximation, the probability that
the
$[R+1-r]$-th strategy is anticorrelated to the $r$-th strategy.
The variance can hence be written analytically as 
\begin{equation}
\sigma_{\rm an}^2 =
\frac{1}{4} \sum_{r=1}^{r=\frac{1}{2}(R-g)} [y_r - P(p)
y_{R+1-r}]^2 +
\frac{1}{4} \sum_{r=1}^{r=\frac{1}{2}(R-g)}
[(1-P(p))y_{R+1-r}]^2 + \frac{g}{4} [y_{\frac{R+1}{2}}]^2
\end{equation} where $g=0$ if $R$ is even and $g=1$ if $R$ is
odd.  The first term represents the net effect after pairing
off the agents  playing anticorrelated strategies.   The second
term in Eq. (3) reintroduces those agents using strategies that
were assumed to be anticorrelated to some more successful
strategy, and hence were discarded unnecessarily in the first
term.  The third term in Eq. (3) is due to the volatility of
the group which remains unpaired in the case where the number of
different strategies used in the calculation is odd. The third
term is usually neglibible compared to the first two.

Figure 1 shows the volatility for $s=2$, $s=4$ and $s=6$ with
$N=101$ agents, as calculated using the analytic results above.
Note that since $m$ is integer, the curves are not smooth. 
Figure 2 compares these theoretical volatility curves (solid
lines) with the numerical simulations (dashed line). The
agreement between the analytic and numerical results is good
across a wide range of $m$ and $s$  values.  In particular, the
analytic results capture the deepening of the minimum in the
volatility as
$s$ decreases.  For the range of $s$ considered, the crowds are
much larger than the anticrowds for small $m$  (i.e., below the
minimum).  Hence the volatility is large for small $m$.  As $m$
increases, the crowds and anticrowds begin to compete
effectively for small
$s$, hence yielding the minimum as discussed qualitatively
earlier in the paper. This minimum disappears with increasing
$s$ since the anticrowds become negligible in size; this is
reasonable since for large
$s$ the likelihood of being stuck with a strategy which is
essentially anti-correlated to a winning strategy is very small.
The agent with such a strategy would typically have
several better choices among his
$s$ strategies.   For large
$m$ (i.e., above the minimum) the crowds and anticrowds have
reduced in size to such a point that the agents act
independently.   The agreement in Fig. 2 can be improved upon by
using a better approximation for $P(p)$ at the expense of
increased analytic complexity.  We note that the above arguments
also explain the  behaviour of the volatility at fixed, 
small $m$ shown in Fig. 1:  the volatility increases
with $s$ due to the decreasing likelihood of 
forming substantial anticrowds.

It is interesting to note that the present analytic approach
uses an idea which is very common in condensed matter physics:
the population of
$N$ agents is treated in terms of  clusters. These clusters are
chosen such that the dominant correlations in the problem
become intra-cluster correlations, i.e., we choose a given
cluster
$i$ to contain the $n_i$ agents using strategy $i$ and the
$n_{\bar i}$ agents using strategy $\bar i$. The strong
intra-cluster correlations between the crowd $n_i$ and anticrowd
$n_{\bar i}$ are treated as accurately as possible, and
dominate the weak inter-cluster correlations which can
themselves be  effectively  ignored.

Finally we note that we have also found a new scaling result for
the minority model for arbitrary $N$, $m$ and $s$, which goes
beyond that reported in Ref. \cite{Savit}. In particular, if we
plot the numerical values of the reduced variable
$\sigma/\sqrt N$ as a function of
$(2 \cdot  2^m)/(Ns)$, then the data undergo almost complete
data collapse \cite{Hui}. This collapse becomes increasingly
precise for larger
$s$ (e.g. $s\geq 4$).
\vskip\baselineskip

\noindent {\bf 4. Bar-Attendance Model}

Arthur \cite{Arthur} proposed the  `bar-attendance' model in
which $N$ adaptive agents, each possessing $s$ prediction rules
or `predictors' chosen randomly from a pool of $V$, attempt to
attend a bar, whose cut-off is
$L$, on a particular night each week. Each week the agents
update their best rule for predicting a given week's attendance
based on the past attendance time-series $x(t)$, which is made
known to all agents.  As stated in Ref.
\cite{Challet}, the minority model seems to be a special case of
the bar-attendance model.  There are differences however: the
output is no longer binary, nor do we restrict all predictors
to depend on the same number of past weeks' data.  The
predictors in our pool of $V$ are chosen from a variety of
`classes' of rules.  For example, one class might comprise
rules which take an arithmetic, geometric,  or weighted average
over the past $m$ weeks' attendances; another class might copy
the result from week $m'$; or alternatively, might take the
mirror image of
$x(t)$ about
$L$ from week $m''$.   In Ref. \cite{Us}, we presented the
results of numerical simulations on the bar-attendance model
and showed that the volatility of the attendance time-series
can exhibit a minimum at small, but finite, $s$. Here we are
interested in the extent to which crowding effects can explain
the dynamics. 

Figure 3 shows a plot of the volatility (i.e., standard
deviation)
$\sigma$ as a function of the number of agents $N$ for a
predictor pool size $V=60$. When the number of agents is smaller
than the cut-off value (i.e. 
$N<L$), the volatility $\sigma \approx 0$ as expected. When the
number of agents is larger than the cut-off value (i.e. 
$N>L$), the volatility increases with increasing $N$
\cite{Us}.  As for the minority model above, suppose that each
agent decides whether to attend based on a pre-assigned
probability. Given the common knowledge of the cut-off
$L$, each agent should attend with a probability $\frac{L}{N}$
and stay away with a probability $1-\frac{L}{N}$. The
volatility can be estimated using the expression for a bounded
random walk,  i.e., the volatility 
$\sigma\sim\sqrt{ N  \cdot \frac{L}{N}  \cdot 
(1-\frac{L}{N})}$ which gives
$\sqrt{L(1-\frac{L}{N})}$. The dashed line in Fig. 3 shows that 
this is however a poor approximation to the numerical results.
The discrepencies imply that we  should include correlations
between the actions of the $N$ agents. In particular several
agents may be using the same predictor at any one time.  Hence,
crowd effects will again be important.  A crowd model can be
developed in a way analogous to that for the minority model. 
However the model will necessarily be less sophisticated than
in Sec. 3 since the `predictor-space' in the
bar-attendance problem is less easily described than the
strategy space for the (binary) minority model. 

Consider the most popular predictor in use at a given stage of
the simulation.  There will be approximately
$n_{1} = N  {\frac{s}{V}}$ agents using this predictor. We
denote these $n_1$ agents as belonging to group $1$. All
$n_1$ agents in group $1$ will make the same decision as whether
to attend in the subsequent week. Note that in the case that
agents, when initially picking predictors from the pool $V$, are
prohibited from picking the same predictor twice, this
expression
$n_1=N {\frac{s}{V}}$ is quite accurate in practice.  When
agents are allowed to pick the same rule repeatedly,  the
expression
${\frac{s}{V}}$ is merely an estimate of $n_1$.  In what
follows, we will disallow repeated-picking of predictors. It is
straightforward to show that the
$r$-th most popular predictor will be used by approximately 
\begin{equation} n_{r}=N {\frac{s}{V}}
\bigg[1-{\frac{s}{V}}\bigg]^{r-1}
\end{equation} agents. These agents belonging to group
$r$ form a crowd and will all make the same decision as to
whether to attend in the subsequent week. As with the minority
model, the resulting volatility-squared $\sigma^2$ is given by
the sum over the volatilities-squared
$\sigma_r^2$ produced by each group, i.e., $\sigma^2=\sum_r
\sigma_r^2$. Now
$\sigma_r^2\sim n_r^2 {\frac{L}{N}}  (1-\frac{L}{N})$, hence
$\sigma$ can be calculated by obtaining
$n_r$ from Eq. (4) above.  Note that $n_r$ is again  rounded to
the nearest integer due to the discreteness of agent-number. 
Figure 3 compares the volatility $\sigma$ obtained using this
crowd model (dotted line) to the numerical simulation (solid
line). The analytic model is in good quantitative agreement
with the simulation results, in stark contrast to the earlier
random walk model without crowding (dashed line).  Future work
will investigate the role of anticrowds in the bar attendance
model.  

\vskip0.2in

\vskip0.5cm
\noindent {\bf 5. Summary}

In summary we have presented an analytic analysis of crowding
effects in the minority and bar-attendance models. We hope that
the present results will stimulate further interest in what is
proving to be an exciting field of study for physicists.   

\vskip0.3in
\noindent {\bf Acknowledgments}

We thank Rob Jonson and C.W. Tai for assistance in the earlier
stages of this work. This work was supported in part (NFJ and
PMH) by a grant from the British Council and the Research
Grants Council of  the Hong Kong SAR Government through the
UK-HK Joint Research Scheme 1998.

\newpage \centerline{\bf Figure Captions}

\bigskip

\noindent Figure 1: Analytic results for the minority model
volatility (i.e., standard deviation) $\sigma$ for
$s=2$,
$s=4$ and
$s=6$ strategies per agent. Number of agents
$N=101$. Since $m$ can strictly only take integer values, the
curves are not smooth.
\bigskip 

\noindent Figure 2: Comparison between analytic and numerical
results for the minority model volatility (i.e., standard
deviation)
$\sigma$. Results are shown for   (a) $s=2$, (b) $s=4$ and (c)
$s=6$, where $s$ is the number of strategies per agent. Number
of agents
$N=101$. Solid line: analytic result (see text). Dashed line:
numerical simulations. Since $m$ is integer, the curves are not
smooth.
\bigskip

\noindent Figure 3: Bar-attendance model volatility (i.e.
standard deviation) 
$\sigma$  as a function of the number of agents $N$. The number
of predictors per agent
$s=3$, the bar cut-off value
$L=60$ and the predictor pool size
$V=60$.  Solid line: numerical simulation results. Dashed line:
random walk model without crowding (see text). Dotted line: 
analytic crowd  model (see text).

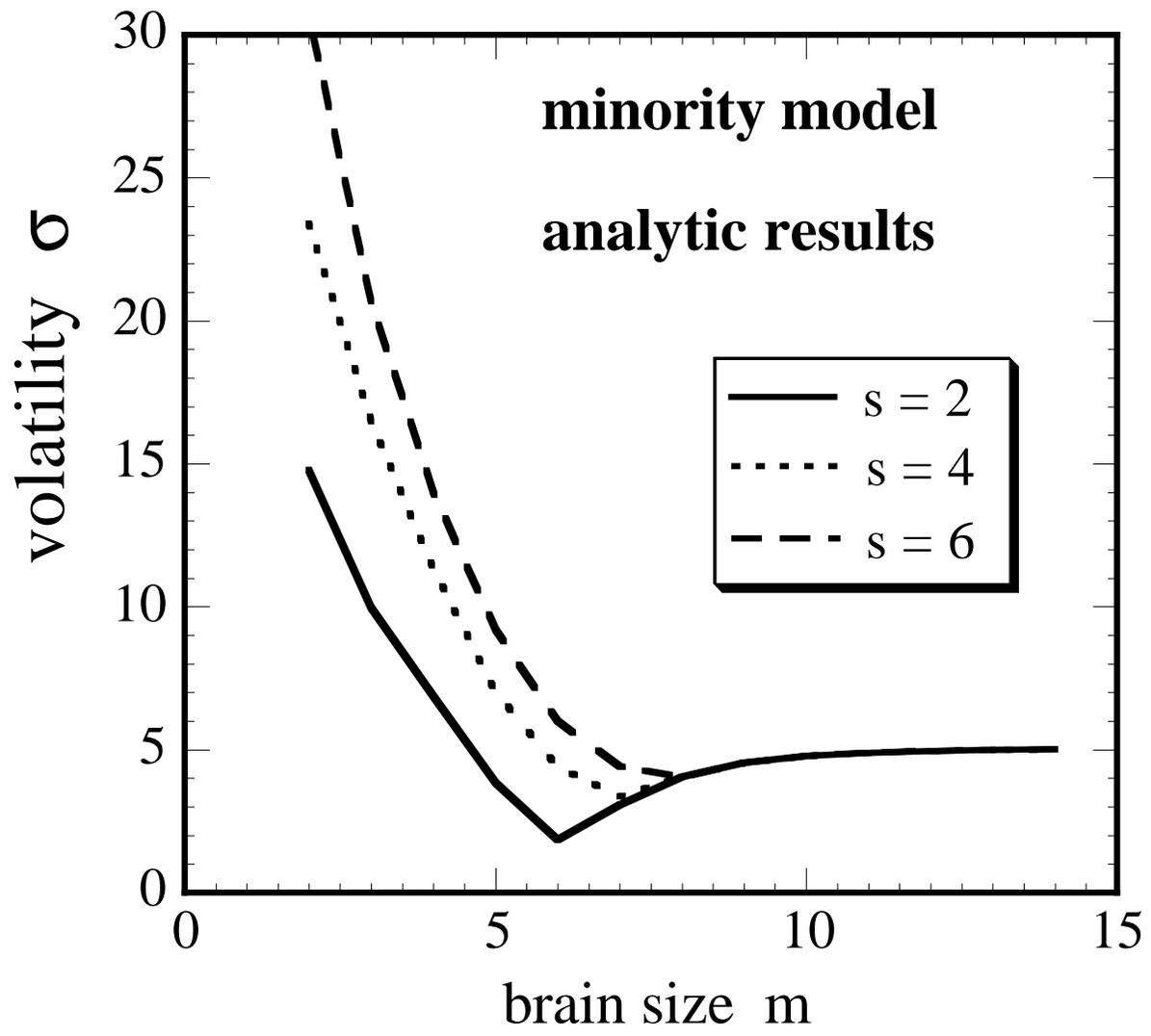
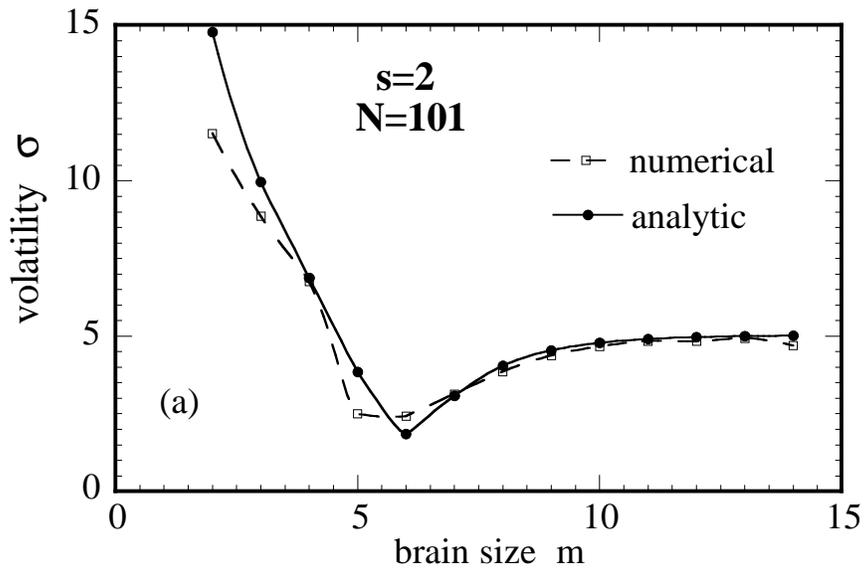
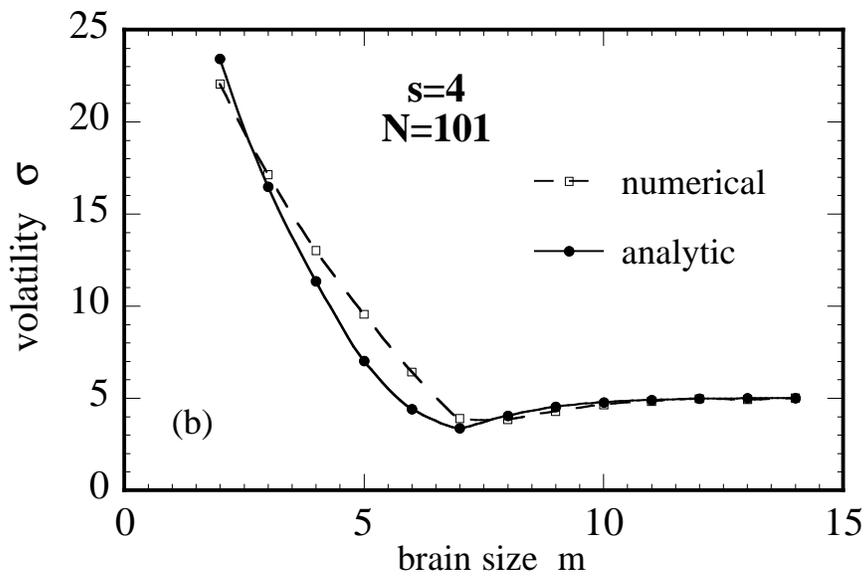
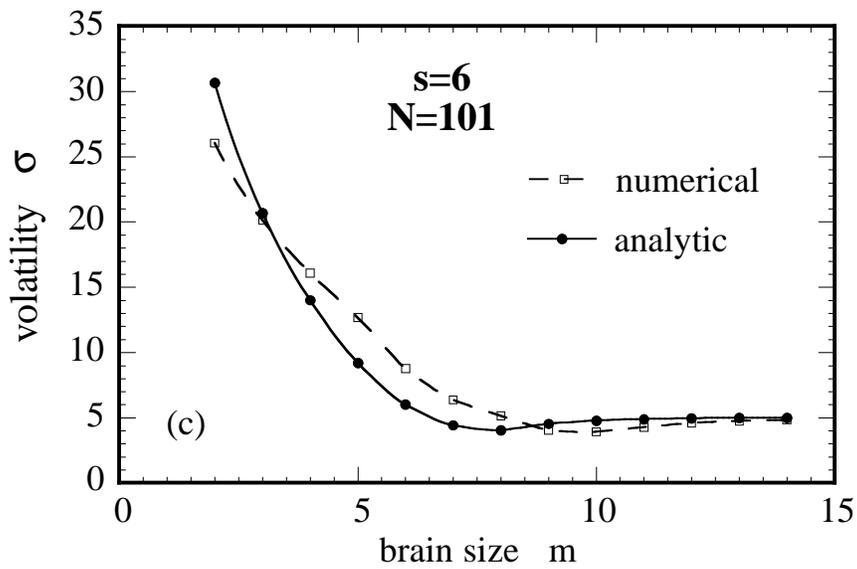
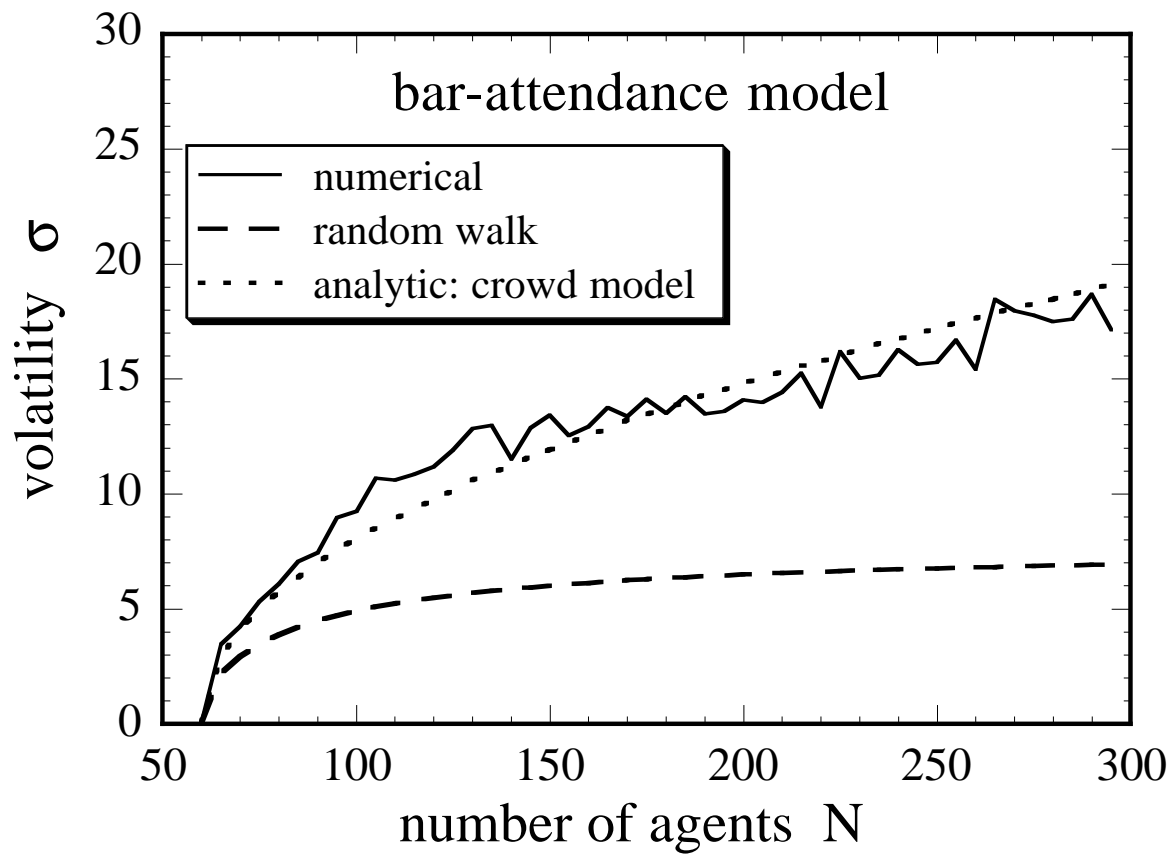
\end{document}